\documentstyle[12pt]{article}
\input epsf
\begin{document}
%\twocolumn 
%\topmargin -1.0truecm
\title{ \hfill
hep-ph/9811252\\ 
\hfill PRL-TH-98/011\\ \vskip 1.5truecm
{\large{\bf Vacuum Solutions of Neutrino Anomalies }}\\
{\large{\bf Through a Softly Broken $U(1)$ symmetry}}} \vskip 2.0truecm

\author{ Anjan S. Joshipura and Saurabh D. Rindani\\
{\ns\it Theoretical Physics Group, Physical Research Laboratory,}\\
{\ns\it Navarangpura, Ahmedabad, 380 009, India.}}

\date{}
%------------------------------------------------------------
\def\ap#1#2#3{           {\it Ann. Phys. (NY) }{\bf #1} (19#2) #3}
\def\arnps#1#2#3{        {\it Ann. Rev. Nucl. Part. Sci. }{\bf #1} (19#2) #3}
\def\cnpp#1#2#3{        {\it Comm. Nucl. Part. Phys. }{\bf #1} (19#2) #3}
\def\apj#1#2#3{          {\it Astrophys. J. }{\bf #1} (19#2) #3}
\def\asr#1#2#3{          {\it Astrophys. Space Rev. }{\bf #1} (19#2) #3}
\def\ass#1#2#3{          {\it Astrophys. Space Sci. }{\bf #1} (19#2) #3}

\def\apjl#1#2#3{         {\it Astrophys. J. Lett. }{\bf #1} (19#2) #3}
\def\ass#1#2#3{          {\it Astrophys. Space Sci. }{\bf #1} (19#2) #3}
\def\jel#1#2#3{         {\it Journal Europhys. Lett. }{\bf #1} (19#2) #3}

\def\ib#1#2#3{           {\it ibid. }{\bf #1} (19#2) #3}
\def\nat#1#2#3{          {\it Nature }{\bf #1} (19#2) #3}
\def\nps#1#2#3{          {\it Nucl. Phys. B (Proc. Suppl.) }
                         {\bf #1} (19#2) #3} 
\def\np#1#2#3{           {\it Nucl. Phys. }{\bf #1}, #3 (19#2)}
\def\pl#1#2#3{           {\it Phys. Lett. }{\bf #1}, #3 (19#2)}
\def\pr#1#2#3{           {\it Phys. Rev. }{\bf #1}, #3 (19#2) }
\def\prep#1#2#3{         {\it Phys. Rep. }{\bf #1} (19#2) #3}
\def\prl#1#2#3{          {\it Phys. Rev. Lett. }{\bf #1} (19#2) #3}
\def\pw#1#2#3{          {\it Particle World }{\bf #1} (19#2) #3}
\def\ptp#1#2#3{          {\it Prog. Theor. Phys. }{\bf #1} (19#2) #3}
\def\jppnp#1#2#3{         {\it J. Prog. Part. Nucl. Phys. }{\bf #1} (19#2) #3}

\def\rpp#1#2#3{         {\it Rep. on Prog. in Phys. }{\bf #1} (19#2) #3}
\def\ptps#1#2#3{         {\it Prog. Theor. Phys. Suppl. }{\bf #1} (19#2) #3}
\def\rmp#1#2#3{          {\it Rev. Mod. Phys. }{\bf #1} (19#2) #3}
\def\zp#1#2#3{           {\it Zeit. fur Physik }{\bf #1} (19#2) #3}
\def\fp#1#2#3{           {\it Fortschr. Phys. }{\bf #1} (19#2) #3}
\def\Zp#1#2#3{           {\it Z. Physik }{\bf #1} (19#2) #3}
\def\Sci#1#2#3{          {\it Science }{\bf #1} (19#2) #3}
\def\n.c.#1#2#3{         {\it Nuovo Cim. }{\bf #1} (19#2) #3}
\def\r.n.c.#1#2#3{       {\it Riv. del Nuovo Cim. }{\bf #1} (19#2) #3}
\def\sjnp#1#2#3{         {\it Sov. J. Nucl. Phys. }{\bf #1} (19#2) #3}
\def\yf#1#2#3{           {\it Yad. Fiz. }{\bf #1} (19#2) #3}
\def\zetf#1#2#3{         {\it Z. Eksp. Teor. Fiz. }{\bf #1} (19#2) #3}
\def\zetfpr#1#2#3{         {\it Z. Eksp. Teor. Fiz. Pisma. Red. }{\bf #1} (19#2) #3}
\def\jetp#1#2#3{         {\it JETP }{\bf #1} (19#2) #3}
\def\mpl#1#2#3{          {\it Mod. Phys. Lett. }{\bf #1} (19#2) #3}
\def\ufn#1#2#3{          {\it Usp. Fiz. Naut. }{\bf #1} (19#2) #3}
\def\sp#1#2#3{           {\it Sov. Phys.-Usp.}{\bf #1} (19#2) #3}
\def\ppnp#1#2#3{           {\it Prog. Part. Nucl. Phys. }{\bf #1} (19#2) #3}
\def\cnpp#1#2#3{           {\it Comm. Nucl. Part. Phys. }{\bf #1} (19#2) #3}
\def\ijmp#1#2#3{           {\it Int. J. Mod. Phys. }{\bf #1} (19#2) #3}
\def\ic#1#2#3{           {\it Investigaci\'on y Ciencia }{\bf #1} (19#2) #3}
\def\tp{these proceedings}
\def\pc{private communication}
\def\ip{in preparation}
\relax

\newcommand{\GeV}{\,{\rm GeV}}
\newcommand{\MeV}{\,{\rm MeV}}
\newcommand{\keV}{\,{\rm keV}}
\newcommand{\eV}{\,{\rm eV}}
\newcommand{\Tr}{{\rm Tr}\!}
\renewcommand{\arraystretch}{1.2}
\newcommand{\beq}{\begin{equation}}
\newcommand{\eeq}{\end{equation}}
\newcommand{\beqa}{\begin{eqnarray}}
\newcommand{\eeqa}{\end{eqnarray}}
\newcommand{\ba}{\begin{array}}
\newcommand{\ea}{\end{array}}
\newcommand{\bmat}{\left(\ba}
\newcommand{\emat}{\ea\right)}
\newcommand{\refs}[1]{(\ref{#1})}
\newcommand{\ler}{\stackrel{\scriptstyle <}{\scriptstyle\sim}}
\newcommand{\ger}{\stackrel{\scriptstyle >}{\scriptstyle\sim}}
\newcommand{\lag}{\langle}
\newcommand{\rag}{\rangle}
\newcommand{\ns}{\normalsize}
\newcommand{\cm}{{\cal M}}
\newcommand{\gr}{m_{3/2}}
\newcommand{\p}{\partial}

\def\rp{ $R_P$} 
\def\321{$SU(3)\times SU(2)\times U(1)$}
\def\tl{{\tilde{l}}}
\def\tL{{\tilde{L}}}
\def\bd{{\overline{d}}}
\def\tL{{\tilde{L}}}
\def\a{\alpha}
\def\b{\beta}
\def\g{\gamma}
\def\c{\chi}
\def\d{\delta}
\def\D{\Delta}
\def\db{{\overline{\delta}}}
\def\Db{{\overline{\Delta}}}
\def\e{\epsilon}
\def\l{\lambda}
\def\n{\nu}
\def\m{\mu}
\def\nt{{\tilde{\nu}}}
\def\p{\phi}
\def\P{\Phi}
\def\x{\xi}
\def\r{\rho}
\def\s{\sigma}
\def\t{\tau}
\def\th{\theta}
\def\ne{\nu_e}
\def\nm{\nu_{\mu}}
\def\rp{$R_P$}
\def\mp{$M_P$}
\def\mgut{M_{HUT}}
\def\emt{$L_e-L_\m-L_{\tau}$ }     
\renewcommand{\Huge}{\Large}
\renewcommand{\LARGE}{\Large}
\renewcommand{\Large}{\large}
\maketitle
\vskip 2.0truecm
\begin{abstract}
We discuss an extended $SU(2)\times U(1)$ model which naturally leads to
mass scales and mixing angles relevant for understanding both the solar and
atmospheric neutrino anomalies in terms of the vacuum oscillations of 
the three known neutrinos. The model uses a softly broken \emt symmetry
and contains a heavy scale $M_H\sim 10^{15}\GeV$. The \emt symmetric
neutrino masses solve the atmospheric neutrino anomaly while breaking of
\emt generates highly suppressed radiative mass scale $\D_S\sim
10^{-10}\eV^2$ needed for the vacuum solution of the solar neutrino
problem. All the neutrino masses in the model are inversely related to
$M_H$, thus providing seesaw-type of masses without invoking any heavy
right-handed neutrinos. Possible embedding of the model into  
an $SU(5)$
grand unified theory is discussed. 
\end{abstract}
\newpage

Recent results on the oscillations of the muon neutrino seen at the
Superkamioka \cite{sk} may be taken as the first experimental evidence for
physics beyond the standard electroweak model. It is attractive to suppose
that these are indirect hints into grand unification.
The neutrino
mass in the  \321 theory can be characterized by a five
dimensional operator which leads to $m_\n\sim {\lag\phi\rag^2\over M}$,
$\lag\phi\rag
\sim 250\GeV$  
being the electroweak  and the $M$ some heavy scale. The
identification of $M$ with a  scale $M_H\sim
10^{15}\GeV$ in grand unified theory nicely fits in \cite{as,wil} with
the neutrino mass
scale $\sqrt{\D_A}\sim 0.07\eV$ seen at the Superkamioka. 

The seesaw model based on grand unified $SO(10)$ theory leads to the above
dimension five term in which $M$ is determined by the right-handed
neutrino masses. Apart from providing an overall scale, this model  also
relates \cite{pm} hierarchy among neutrino masses to that in the masses of 
the other (up quarks in
the minimal case) fermions. This feature of $SO(10)$ can indeed provide
another scale $\D_S$ needed to solve the solar neutrino problem. In the
simplest $SO(10)$ model one expects ${\D_S\over \D_A}\sim \left({m_c\over
m_t}\right)^4$.
$\Delta_A\sim 10^{-3}\eV^2$ then automatically leads to a $\D_S$ required
for the vacuum solution \cite{vac} to the solar neutrino problem. The two
large mixing
angles needed in this case are not generic features of the seesaw model
but could  come out under reasonable assumptions. \cite{bhs,yn}. 

The above attractive features of $SO(10)$ related to neutrino masses
are not shared by generic $SU(5)$-based grand unified models. It is
possible in these models to obtain neutrino masses and also to understand
their overall scale  in terms of the grand unified
scale simply by adding a heavy 15 -dimensional Higgs field
\cite{as,pm,ms} . But one cannot
easily relate hierarchy in $\Delta_S$ and $\Delta_A$ to the known fermion
masses as in the $SO(10)$ case. Our aim here is to present a simple
$SU(5)$ scheme which does this. While the mechanism we discuss is more
general, 
we give a specific example in which (a) ${\D_S\over\D_A}$ gets related to
the charged lepton masses and (b) two large mixing angles come out
naturally. The natural value for the ${\D_S\over\D_A}$ is close to
$10^{-7}$ resulting in vacuum oscillations as the cause for both the solar and
atmospheric neutrino deficits.

To simplify the matter we shall first discuss a scheme based on the
standard $SU(2)\times U(1)$ model and discuss its $SU(5)$ generalization
later on. We need to extend $SU(2)\times U(1)$ model in two ways.
We enlarge it with two extra multiplets of scalar fields namely a triplet
$\Delta$ and an additional doublet field $\phi_2$. We also impose a global
\emt symmetry. This symmetry has been recognized \cite{as,bhs} to provide
under
reasonable assumptions two large
mixing angles needed for the vacuum solutions of the neutrino anomalies.
It leads to a pair of degenerate neutrinos with a common mass $m_0$ which
determine the atmospheric  neutrino mass scale. $m_0$ is inversely related
to the grand unified scale $M_H$ in the manner discussed below.

Keeping $SU(5)$ unification in mind, we assume the triplet to 
be very heavy, with mass $\sim M_H$. But such a heavy triplet can
influence the low energy theory crucially  by generating \emt symmetric
neutrino
mass matrix at tree
level and
departure from it at one-loop level.

The leptonic Yukawa couplings in the model are given by
\beq \label{yukawa1}
-{\cal L}_Y={1\over 2} f_{ij}\bar{l}_{iL}^{c\prime}\;\D\;l_{jL}' +
\Gamma^a_{ij}\bar{l}_{iL}'e_{jR}'\phi^a +\;\; H.c.\;,\eeq
where $a=1,2$ label the Higgs doublets and $\D$ is $2\times 2$ matrix in
the $SU(2)$ space.
The \emt symmetry allows the following Yukawa textures:
\beqa \label{yukawa2}
f&\equiv& {M^\n_0\over \lag\Delta^0\rag}= {m_0\over \lag\Delta^0\rag}
\left(
\ba{ccc}
0&c&s\\
c&0&0\\
s&0&0\\
\end{array}\right)\;\;;\nonumber \\
\Gamma_1&\equiv& {M^l_1\over \lag\phi_1^0\rag}={1\over
\lag\phi_1^0\rag}\left(
\ba{ccc}
m_{1}&0&0\\0&m_2&m_{23}\\0&m_{32}&m_3\\ \ea
\right)\;\;; \nonumber\\
\Gamma_2&\equiv&{M_2^l\over \lag\phi_2^0\rag}={1\over
\lag\phi_2^0\rag}\left(
\ba{ccc}
0&m_{12}&m_{13}\\0&0&0\\0&0&0\\ \ea
\right)\; ,
\eeqa
where we have chosen the \emt charge 2 for the field $\phi_2$ and zero
for $\phi_1$ and $\Delta$.

The tree level neutrino mass matrix is \emt symmetric and can be diagonalized by
\beq\label{u}
U^\n=\left( \ba{ccc}
1/\sqrt{2}&-1/\sqrt{2}&0\\
c/\sqrt{2}&c/\sqrt{2}&-s\\
s/\sqrt{2}&s/\sqrt{2}&c\\
\end{array}\right)\;\; .
\eeq
If mixing among charged leptons is small then $U^\n$ provides the
bimaximal mixing \cite{bim} for $c\sim s\sim {1\over 
\sqrt{2}}$ and can therefore simultaneously solve the solar and
atmospheric neutrino anomaly
through the vacuum oscillations.

The atmospheric scale $m_0$ is determined in the model by the vacuum 
expectation value
(vev) of $\Delta^0$. This is driven  by the following   scalar
potential  
\beqa \label{pot}
V&=&M_a^2 \phi^{\dagger}_a \phi_a+M_{H}^2 Tr.
\Delta^{\dagger}\Delta \nonumber \\
&+&\lambda_a (\phi_a^{\dagger} \phi_a)^2+\lambda_{\Delta} Tr.
(\Delta^{\dagger}\Delta)^2+....\nonumber
\\
&-&\left[\m_{ab} \phi_a^T\Delta\phi_b+c.c.\right]\; . 
\eeqa
The terms not explicitly written in the above equations correspond to
some of the quartic terms involving $\Delta$ and quartic cross terms 
for the doublet fields. The trilinear terms in eq.(\ref{pot}) 
are of crucial importance. Firstly, they induce a small vev for the 
neutral Higgs $\Delta^0$ leading to a degenerate pair of neutrinos. 
In addition,
they softly break the lepton number and \emt
symmetry. This breaking makes the model
phenomenologically acceptable which otherwise would have contained a
doublet plus triplet majoron already ruled out at LEP. In addition,
the \emt breaking by  trilinear terms also generate radiative corrections
to the neutrino mass matrix which result in the splitting of the 
degenerate pairs and solves the solar neutrino problem.

The
triplet vev following from
eq.(\ref{pot}) after minimization is of the order
\beq 
\lag\Delta^0\rag\sim {\lag\phi_1\rag \lag\phi_2\rag\over M_{H}} \;
,\eeq
where $ \m_{ab}$  are assumed to be of the same order as the
(large) triplet mass $M_{H}$. The neutrino mass generated at tree
level thus displays the seesaw type dependence on the heavy scale.
Specifically, one gets through eq.(\ref{yukawa2}) 
$m_0\sim 3 (10^{-1}-10^{-2}) \eV$ for
$m_{H}\sim
10^{14}-10^{15}\GeV$ and $ \lag\phi_1\rag\sim \lag\phi_2\rag$ providing
the
atmospheric neutrino scale.

The tree level neutrino mass matrix following from  eq.(\ref{yukawa2})
is \emt symmetric but the presence of a vev for $\phi_2$ breaks this
symmetry
in the charged lepton mass matrix. This breaking ultimately gets communicated 
to the neutrino mass matrix at the one-loop level. This occurs through the one
loop diagrams shown in Fig.(1). 

Let us define the charge lepton mass eigenstates as $e_{i L,R}\equiv
U^{\dagger L,R}_{i\a}e_{\a L,R}'$   where
\beq \label {ul}
U^{L\dagger} (M^l_1+M^l_2)U^{R}\equiv U^{L\dagger} M^l U^{R}=M_0^l\;\; ,
\eeq
$M_0^l$ being the diagonal charged lepton mass matrix. 
The soft breaking
of \emt through $\m_{22,23}$ and vev for $\phi_2$ results in  finite and
calculable corrections to
\emt breaking entries of the neutrino mass matrix $M_\n$. In order to 
evaluate these, it is
convenient
to work with the original (massless) neutrino flavor basis and treat the
mass term $M_0^\nu$ as an additional interaction. The $H_a$ in Fig.1 refer to the 
mass eigenstates of the charged Higgs fields $H'_a\equiv
(\phi_1^+,\phi^+_2,\Delta^+)=O_{ab}H_b$. 

We have evaluated  diagrams of Fig. 1 in the $R_{\x}$ gauge. Each of the
diagrams 
gives a finite correction to the \emt breaking elements in $M^\n_0$ and their
sum is gauge independent. One finds,

\beqa \label{main}
(M^\n)_{11}&=&{g^2\over 16\pi^2  M_W^2}
 (M^\n_0U^{ L}M_0^lM_0^{l\dagger}U^{\dagger L})_{11}\times \nonumber
\\
&\;& \left(
1-3ln{M_W\over M_3} - {4 M_W^2 O_{22}\over g^2 \lag\phi_2^0\rag}\left(
{\sqrt{2}O_{32}\over \lag\Delta^0\rag}-{O_{12}\over2
\lag\phi^0_1\rag}\right)
ln{M_2\over M_3}\right) \nonumber \\
(M^\n)_{ij}&=&{g^2\over 32\pi^2 M_W^2}\left(
(M^\n_0U^{L}M_0^lM_0^{l\dagger}U^{\dagger L})_{ij}+
(M^\n_0U^{ L}M_0^lM_0^{l\dagger}U^{\dagger L})_{ji}\right)\times \nonumber
\\ 
&\;& \left(
1-3ln{M_W\over M_3}- {4 M_W^2O_{12}\over g^2\lag\phi_1^0\rag}\left(
{\sqrt{2}O_{32}\over \lag\Delta^0\rag}-{O_{22}\over2
\lag\phi^0_2\rag}\right)
ln{M_2\over M_3} \right)\; .
\eeqa
$i,j$ in the above equation take the value 2 and 3 only. $M_0^l$ 
and $U^L$ are defined in eq.(\ref{ul}). We have
repeatedly used the orthogonality of the matrices $U^{L,R}$ and $O$ in
arriving at finite result. $M_{2,3}$ refer to the masses of the two
physical charged Higgs fields one of which is very heavy, i.e. $M_3\sim
M_{H}$. Terms  cubic in neutrino masses are neglected in 
writing the above results.

Although the heavy field decouples in the limit $M_H$ very large,
its
residual mixing of order ${  M_W\over M_{H}}\sim {m_0\over
M_W}$
with the doublet
fields  influences the radiative masses. This is explicit in the above
equations through the presence of the tree level neutrino mass matrix.
This has the consequence that  the radiatively generated mass terms also
display the basic seesaw
structure present at the tree level.

The contributions in eq.(\ref{main}) depend on all three
charged lepton masses but the contribution due to tau lepton
dominates over the rest unless $U^L_{31}$ is enormously
suppressed . We shall assume dominance of this
contribution. The (logarithmic)
contribution of the $W$ diagram is similar in magnitude
to the Higgs contributions containing elements of $O$ if the mixing among
doublet fields $\phi_{1,2}$
is O(1). Hence for the numerical estimate we shall concentrate on the 
$ln {M_W\over M_3}$ term. The radiatively corrected neutrino mass
matrix 
then has the structure
\beq \label{app} M^\n\approx m_0\left( \ba{ccc}
2\e s&c&s\\
c&0&\e c\\
s&\e c&2\e s\\ \ea \right)\; . \eeq
We have implicitly assumed a real $U_L$ and $U^L_{33}\gg U^L_{23}$ in
writing the above
structure.  The parameter $\e$
is defined as
\beq
\e\equiv -{3 g^2m_\t^2\over 32 M_W^2\pi^2} ln{M_W\over
M_3} U^L_{13}U^L_{33}\sim (7 \times 10^{-5}) U^L_{13}U^L_{33}\eeq
when $M_3\sim 10^{15}\GeV$.

Let us now look at the phenomenological consequences. As already
mentioned, $\Delta_A\equiv m_0^2\sim 10^{-2}-10^{-3} \eV^2$ follow
when Higgs mass  $M_{H}$ is in the range $10^{14}-10^{15}\GeV$.
The radiatively corrected mass matrix also implies:
\beq \label{ratio}
{\Delta_S\over \D_A}\sim 8 \e s\sim ( 4\times 10^{-4})U^L_{31}U^L_{33}\leq
2\times 10^{-4}.
\eeq
The mixing among neutrinos is governed by 
\beq \label{km}
K_l\equiv U_{L}^{\dagger} U_\n \; .\eeq

The ratio ${\Delta_S\over \D_A}$ depends upon unknown values of the
mixing among charged
leptons. The scale required for the vacuum solution follows 
if the mixing element $U_{31}$ is small. Indeed, $U_{ij}\sim
U_{ji}\sim O({
m_i\over m_{j}})$ , for $i<j$ leads to
$$ \D_S\sim  10^{-7} \D_A\sim 10^{-9}-10^{-10} \eV^2 \; .$$
The leptonic Kobayashi-Maskawa matrix $K$ is also approximately
given in this case by $U^\n$ which provides the required bimaximal mixing.
Thus model under consideration leads to
vacuum solution to the solar neutrino problem for natural value of the
relevant parameters.

Unlike the vacuum case, the MSW \cite{msw} solution does not follow
naturally in the model. To see this, let us concentrate on the 
approximate result
eq.(\ref{ratio}). If $U^L_{33}U^L_{31}$ is less than O(1) then
one does not get
a $\Delta_S$ in the range required for the MSW to work inside the Sun
even when $\D_A$ is close to its upper limit of $10^{-2} {\rm eV}^2$. 
Moreover, the charged-lepton mixings being small, the relevant
\cite{sl} effective
mixing angle $\sin^2 2\theta_S\equiv {4 K_{e1}^2K_{e2}^2\over
(1-K_{e3}^2)^2}$ is close to 1  in this case, and one gets energy
independent suppression already ruled out \cite{bsk,giunti} at the 99\%
CL.

On the other hand, if mixing in the charged-lepton sector,
specifically $U^L_{31,33}$, is large, there is a possibility that the large
mixing among the neutrinos can be compensated by  the large mixing among the
charged leptons. The effective mixing angle in that case can be 
appreciably less than 45$^\circ$. Recent global fit to new
experimental results does allow
large mixing angle solution if one does not include the Superkamioka
results on the day-night asymmetry in the fit. However, in that case, the
allowed value of $\D_A$ is even smaller than in the small-angle case.
Specifically, the allowed range for large mixing solution is
given by \cite{bsk}
$$ 0.6<\sin^22\theta_S<0.8\;\;\;\;  ; \;\;\;\;8\times
10^{-5}\eV^2<\D_S<2\times
10^{-4}\eV^2. $$

It follows that even though  proper choice of $U^L_{31}$ can lead to the
correct $\sin^2 2\theta_S$, eq.(\ref{ratio}) cannot lead to
the $\D_S$ in the required range. There is the possibility that the Higgs
contribution we have neglected 
might, for some
choice of Higgs and charged-lepton mixing, give rise to $\D_S$ in the allowed
region. However, this would be a marginal case.

The generalization of above results to  $SU(5)$ model is straightforward.
As an illustration, consider a model with a $15$-plet 
$\D$ and two Higgs $\bar{5}$-plets $\phi_{1,2}$. The \emt
symmetry can be replaced by a  $U(1)_H$ symmetry under
which three generations of $\bar{5}$-plet of fermions carry 
charges (1,-1,-1) respectively while the corresponding 10-plets have
opposite $U(1)_H$ charges. 
$\phi_2$ carries charge 2 and
rest of the fields are taken neutral. In this case down quarks
together with the charged lepton have the mass structure given
by $M^l$  while up-quark masses are given by the following Yukawa
couplings:

\beq \label{yukawa3}
-{\cal L}_u=\Gamma^{u}_{a\;ij}10_{i}10_{j}\phi^{*a} \;,\eeq
where $a=1,2$ label the two $\bar{5}$-plets of Higgs.
The \emt symmetry allows the following Yukawa textures:
\beqa \label{yukawa4}
\Gamma^u_{1}&\equiv& \left( 
\ba{ccc}
0&\b_1&\b_2\\\b_1&0&0\\ \b_2&0&0\\ \ea
\right)\;\;\; ; \nonumber\\
\Gamma^u_2&\equiv& \left(
\ba{ccc}
0&0&0\\0&\b_{22}&\b_{23}\\0&\b_{23}&\b_{33}\\ \ea
\right)\; .
\eeqa
It follows that 
the additional $U(1)_H$ symmetry does not lead to any prediction in the
quark sector but allows a general structure for the quark masses and
mixing.

The trilinear
terms in (\ref{pot}) are allowed by  $SU(5)$ but break the $U(1)_H$
softly. All the previous considerations on the tree level as well
radiative neutrino masses go through. However there are additional
diagrams similar to Fig. 1 contributing to the neutrino masses. These are
obtained from above by replacing $W$ boson, charged leptons and colour
singlet Higgs by the heavy charge-1/3 $X$-bosons, d-quarks and the colour
triplet Higgs bosons respectively. The contribution of these is suppressed
due to heavy $X$ mass and due to the fact that colour-triplet Higgs have
comparable masses. This is to be contrasted with Fig. 1 which
contributes
large logarithmic factor due to vastly different Higgs masses in the loop,
see eq. (\ref{main}). Thus previous considerations based on the
$SU(2)\times U(1)$ model remain valid in this case.

We have discussed here a specific case of the \emt symmetry in view of its
phenomenological interest. But the suggested basic mechanism provides a
nice
scheme to generate pseudo-Dirac structure for neutrino masses in which
some symmetry (e.g. $L_e-L_\m$) leads to Dirac structure and its violation
in the charged lepton sector leads to splitting among the degenerate
pair
radiatively.

Since there have been numerous schemes \cite{mb} for radiative neutrino
masses,
it is appropriate to contrast the present one from the rest.
Large class of radiative models \cite{rad} use the original mechanisms
proposed by Zee \cite{zee} and  by Babu \cite{babu}. The violation of
lepton number at tree level
gets communicated radiatively to neutrinos in these schemes.
Here, neutrinos have lepton number violating but \emt symmetric masses
at tree level and breaking of \emt symmetry gets communicated radiatively.

The most noteworthy feature of the present scheme is the dependence of the 
radiative corrections on the tree-level neutrino  and the charged
lepton masses.  The former is absent in Zee type of models and
radiatively generated contribution is controlled only by the charged
lepton masses. This feature makes the radiative contribution here
quite small and allows one to obtain extremely suppressed solar neutrino
scale relevant for the vacuum solution of the solar neutrino problem. 

The
conventional radiative models need introduction of additional
singly and doubly charged Higgs fields with masses near electroweak
scale. Here the role of the charged singlet is played by corresponding
field in the triplet which is very heavy. Thus the present scheme does
not predict  a light exotic charged Higgs. Theoretically, the conventional
models are not easily amenable to grand unification in contrast to
the present case.  The present scheme is similar
in spirit to the seesaw model based on $SO(10)$ . In spite of the
absence of the right-handed neutrino , the model presented here 
contains seesaw structure
for all the neutrino masses and these masses are closely linked to the 
mass of the charged leptons. This makes the model fairly predictive and 
leads to a simultaneous solution for the solar and atmospheric
anomalies which to date provide the strongest hints to believe that
neutrinos are massive.

\newpage
\begin{figure}[t]
\epsfxsize 15 cm
\epsfysize 15 cm
\epsfbox[25 151 585 704]{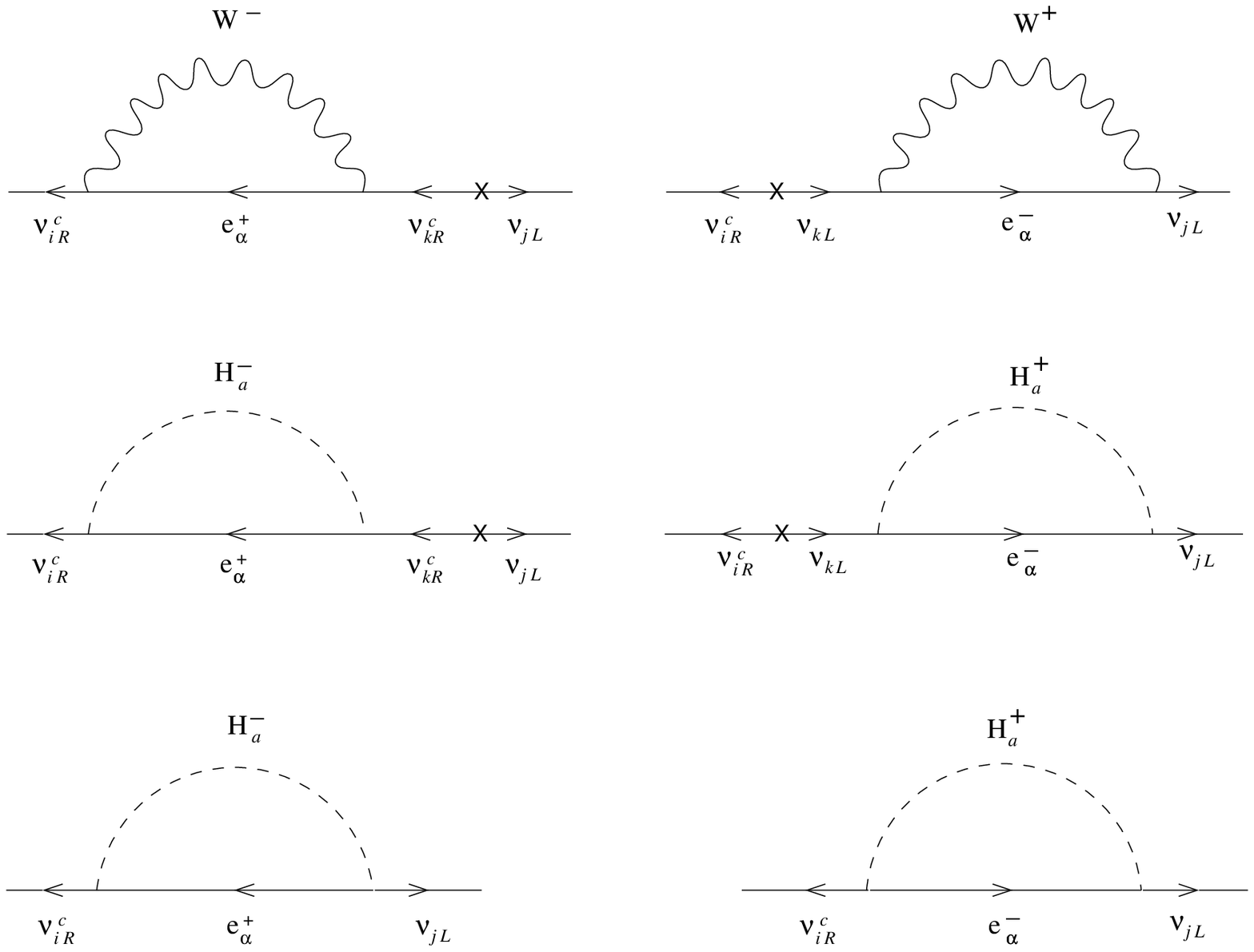}
\label{Fig. 1}
\caption {\sl 1-loop diagrams contributing to the \emt
breaking entries of the neutrino mass matrix.}  
\end{figure}

\end{document}